\documentstyle{article}
\def\be{\begin{equation}}
\def\ee{\end{equation}}
\def\bea{\begin{eqnarray}}
\def\eea{\end{eqnarray}}
\def\l{\label}

\def\r{\ref}

\def\case#1/#2{\textstyle\frac{#1}{#2}}
\def\k0{\kappa_{0}}

\title{$D$-Dimensional Gravity from $(D+1)$ Dimensions}
\vspace{0.6cm}
\normalsize
\vspace{2cm}
\author{Steve Rippl$^{1a}$, Carlos Romero$^{1,2b}$
\& Reza Tavakol$^{1c}$\thanks{e-mail (a) sfr@maths.qmw.ac.uk, (b)
car@maths.qmw.ac.uk, (c) reza@maths.qmw.ac.uk} \\
\\
$^1$ School of Mathematical Sciences\\
Queen Mary \& Westfield College\\
Mile End Road\\London E1 4NS, UK\\
\\
$^2$ Departamento de F\'{\i}sica\\
Universidade Federal da Para\'{\i}ba\\
C. Postal 5008 - J. Pessoa -Pb\\
58059-970 - Brazil}
\vspace{2cm}
\begin{document}
\sloppy
\maketitle

\begin{abstract}
We generalise Wesson's procedure, whereby vacuum $(4+1)-$dimensional
field equations give rise to $(3+1)-$dimensional equations
with sources, to arbitrary dimensions.
We then employ this generalisation to relate the usual $(3+1)-$dimensional
vacuum field equations to $(2+1)-$dimensional field equations with
sources and derive the analogues of the classes of solutions obtained
by Ponce de Leon. This way of viewing lower dimensional gravity theories
can be of importance in establishing
a relationship between such theories and the usual 4-dimensional
general relativity, as well as giving a
way of producing exact solutions in $(2+1)$ dimensions that are
naturally related to the vacuum $(3+1)-$dimensional solutions.
An outcome of this correspondence, regarding the nature of
lower dimensional gravity, is that
the intuitions
obtained in $(3+1)$ dimensions
may not be automatically transportable to lower dimensions.

We also extend a number of physically motivated solutions
studied by Wesson and Ponce de Leon
to $(D+1)$ dimensions and employ the
equivalence between the $(D+1)$ Kaluza-Klein theories with empty
$D-$dimensional Brans-Dicke theories (with $\omega=0$)
to throw some light on the solutions derived by these authors.
\end{abstract}
\newpage
\section{Introduction}
Since the advent of general relativity (GR), the aim of a purely
geometrical description of all physical forces as well as
that of a geometrical origin for the matter content
of the Universe
has been pursued \cite{dav85,ein56,salam80}. A simple and elegant
idea in this connection,
recently put forward by Wesson, Ponce de Leon and Coley
\cite{w92,wpdl92,coley94},
involves starting from the vacuum 5-dimensional Kaluza-Klein
equations
\be
^{(5)}R_{ab} =0,
\l{kk}
\ee
where $R_{ab}$ is the Ricci tensor in the 5-dimensional
space.
They show that equation (1) gives rise
to $4-$dimensional
Einstein equations {\it with sources} in the form
\be
^{(4)}G_{\alpha\beta}= ^{(4)}T_{\alpha\beta},
\l{gr}
\ee
provided the extra terms related to the fifth dimension
are appropriately used to define an energy momentum
tensor $T_{\alpha\beta}$ \cite{wpdl92}.
Furthermore, it has been shown that for certain cosmological
solutions of the vacuum $(4+1)$ field equations (\r{kk}),
the resulting sources in the $(3+1)$ dimensions
can be interpreted as matter with reasonable
equations of state \cite{w92}.
\\

Parallel to these developments, there has recently been a great deal
of interest in lower dimensional gravity theories
\cite{mann93,kuchar84,mann95}.
The usual motivation for this
body of work is that by enormously simplifying the difficulties
usually associated with the Einstein field equations in $(3+1)$ dimensions,
such theories
can act as ``laboratories'' in which models can be more readily
constructed to study a number of phenomena,
ranging from quantum gravitational
effects \cite{odin} and topological defects \cite{deser} to
black holes \cite{mann93,mann95}.
The hope is that these lower dimensional models are of value
in understanding their analogues in $(3+1)$
dimensions.

The problem, however, is that such theories
have radically different properties to the 4-dimensional GR, for example,
the non-existence of a Newtonian limit \cite{kuchar84}.
It is therefore not at all clear that intuitions
obtained from the study of lower dimensional models would be of relevance
in 4-dimensional GR. Similarly, it is not clear that $(3+1)-$dimensional
intuitions are automatically transportable to
$(2+1)$ dimensions.
We shall, for example, see that there exist vacuum Einstein
solutions in $(3+1)$ dimensions
which give rise to solutions in $(2+1)$ dimensions with
corresponding sources, which
on the basis of intuitions imported from ordinary gravity
could be ruled out as ``non-physical''. The real physics, however,
takes place in (at least) $(3+1)$ dimensions and the notion of physicality
of solutions in lower dimensions can only make sense in relation to
the usual $(3+1)$ theory.
An important issue is, therefore, to understand
how lower dimensional theories are related to the usual
4-dimensional theory.
\\

Our aims here are threefold:
\\

\noindent (i) to generalise the scheme put forward by Wesson {\it et al}.
for relating the vacuum $(D+1)-$dimensional theories to $D-$dimensional
theories\footnote{By $D-$dimensional we mean
$((D-1),1)-$dimensional.}
with sources, and generalising some of the physically motivated
solutions to $D$ dimensions,
\\

\noindent (ii) to employ the equivalence between the $(D+1)-$dimensional
Kaluza-Klein theory with the empty Brans-Dicke theory in $D$
dimensions in order to elucidate some of the solutions derived by
Wesson and others, and
\\

\noindent (iii) to employ this scheme as a tool to study
the relation between lower dimensional gravity theories and the
usual 4-dimensional vacuum GR, as well as
a way of producing new solutions in
lower dimensional gravity theories.
%
\section{$D-$dimensional gravity from $(D+1)$ dimensions}
In this section we generalise the work of Wesson and Ponce de Leon
\cite{wpdl92} to $(D+1)$ dimensions. Our motivation for this extension
is twofold. Firstly to justify, mathematically, the analogue
of the procedure by Wesson {\it et al} in going from $(3+1)$ to $(2+1)$
dimensions,
which will be discussed in section (4) below. Secondly,
our aim is to ultimately consider a generalisation of the scheme of
Wesson {\it et al}, which would involve a ``cascade'' of such steps. There
are, however,  important differences here after the first step
and we shall return to these in a future publication.

To do this we start with the $(D+1)-$dimensional
source-free Kaluza-Klein field equations (\r{kk})
and in line with the ansatz adopted in \cite{wpdl92},
we take the metric to be in the form\footnote{
In this section the Latin and Greek indices run from  $0$ to $D$
and $0$ to $(D-1)$ respectively.}

$$\pmatrix{~ &~ &~&~&~&0 \cr
{}~ &~ &~&~&~&0 \cr
{}~ &~ &g_{\alpha\beta} &~&~&0 \cr
{}~ &~ &~&~&~&0 \cr
0 &0 &0&0&~&g_{DD}\cr},$$
 where both the $(D+1)$ and $D-$dimensional
metrics $g_{ab}$ and
$g_{\alpha\beta}$ are in general dependent on the coordinate
$x^D$ and as in \cite{wpdl92} we write the
$g_{DD}$ as
\be
g_{DD} = \phi^2, ~~~~~~~~g^{DD} = \frac{1}{\phi^2}.
\l{gdd}
\ee

Now proceeding as in \cite{wpdl92}
 and starting from the $(D+1)$-dimensional Ricci tensor
expressed in terms of the $(D+1)$ Christoffel symbols
\be
R_{ab} = ({\Gamma^c}_{ab})_{,c} - ({\Gamma^c}_{ac})_{,b}
+{\Gamma^c}_{ab}{\Gamma^d}_{cd} - {\Gamma^c}_{ad}{\Gamma^d}_{bc},
\ee
and letting $a\rightarrow \alpha, ~~b\rightarrow \beta$, we
obtain the $D-$dimensional part of the $(D+1)$-dimensional
Ricci tensor indicated by $^{(D+1)}R_{\alpha \beta}$ thus:
\bea
^{(D+1)}R_{\alpha \beta}&=&^{(D)}R_{\alpha \beta} + ({\Gamma^D}_
{\alpha\beta})_{,D}- ({\Gamma^D}_{\alpha D})_{,\beta} +
{\Gamma^\nu}_{\alpha\beta}{\Gamma^D}_{\nu D} +
{\Gamma^D}_{\alpha\beta}{\Gamma^d}_{Dd}\nonumber \\
&-& {\Gamma^D}_{\alpha\lambda}
{\Gamma^\lambda}_{\beta D}
-{\Gamma^d}_{\alpha D}{\Gamma^D}_{\beta d}.
\l{5}
\eea
Substituting in (\r{5}) from (\r{gdd}), we obtain
\bea
^{(D+1)}R_{\alpha \beta}= ^{(D)}R_{\alpha \beta} - \frac{\phi_{\alpha;\beta}}
{\phi}
&+&\frac{1}{2\phi^2} \left (\frac{\phi_{,D}}{\phi} g_{\alpha\beta , D}
-g_{\alpha\beta ,DD} + g^{\lambda \mu} g_{\alpha \lambda ,D}
g_{\beta \mu , D} \right. \nonumber \\
&-& \left. \frac{1}{2} g^{\mu \nu} g_{\mu\nu , D}
g_{\alpha\beta , D}\vphantom{\frac{\phi_{,D}}{\phi}} \right ),
\l{galbe}
\eea
and similarly we obtain an analogous expression for $^{(D+1)}R_{DD}$,
\bea
^{(D+1)}R_{DD} &=& ({\Gamma^\alpha}_{DD})_{,\alpha} -
({\Gamma^\alpha}_{D\alpha})_{,D} + {\Gamma^\alpha}_{DD}
{\Gamma^\beta}_{\alpha\beta} + {\Gamma^D}_{DD}
{\Gamma^\beta}_{D\beta}\nonumber \\ &-&{\Gamma^\alpha}_{D\beta}
{\Gamma^\beta}_{D\alpha} - {\Gamma^D}_{D\beta}
{\Gamma^\beta}_{DD}.
\eea
Now the source free field equations in $(D+1)$ dimensions are given by
\be
^{(D+1)}R_{ab} = 0 \Longrightarrow ^{(D+1)}R_{\alpha\beta} = 0,~~
^{(D+1)}R_{DD} = 0,~~ ^{(D+1)}R_{\alpha D}= 0.
\l{8}
\ee
 From the first of these ($^{(D+1)}R_{\alpha\beta} = 0$) we obtain (using
(\r{galbe}))
\be
^{(D)}R_{\alpha \beta} = \frac{\phi_{\alpha;\beta}}{\phi}
- \frac{1}{2\phi^2} \left (\frac{\phi_{,D}}{\phi} g_{\alpha\beta , D}
-g_{\alpha\beta ,DD} + g^{\lambda \mu} g_{\alpha \lambda ,D}
g_{\beta \lambda , D} -\frac{1}{2} g^{\mu \nu} g_{\mu\nu , D}
g_{\alpha\beta , D}\right ).
\ee
The second equation ($^{(D+1)}R_{DD} = 0$) results in
\be
\phi \Box{\phi} = -\frac{1}{4} g^{\lambda\beta}_{~~, D} g_{\lambda \beta , D}
 -\frac{1}{2} g^{\lambda\beta } g_{\lambda \beta ,D D}
 +\frac{\phi_{,D}}{2\phi} g^{\lambda\beta } g_{\lambda \beta , D},
\ee
One can now define an effective energy-momentum tensor
by
\bea
^{(D)} T_{\alpha\beta} &=& \frac{\phi_{\alpha;\beta}}{\phi}
- \frac{1}{2\phi^2} \left[\frac{\phi_{,D}}{\phi} g_{\alpha\beta , D}
-g_{\alpha\beta ,D D} + g^{\lambda \mu} g_{\alpha \lambda ,D}
g_{\beta \mu , D} - \frac{1}{2} g^{\mu \nu} g_{\mu\nu , D}
g_{\alpha\beta , D} \right.
\nonumber \\ &+&\frac{ g_{\alpha\beta}}{4}
\left. \left[{g^{\mu\nu}}_{,D} g_{\mu\nu , D}
+ (g^{\mu\nu} g_{\mu\nu , D})^2\right] \vphantom{\frac{\phi_{,D}}
{\phi}} \right],
\l{eq10}
\eea
where
\be
^{(D)}T_{\alpha\beta} = ^{(D)}R_{\alpha\beta} - \frac{1}{2}
^{(D)}R g_{\alpha\beta}.
\ee

In this way the $(D+1)$-dimensional source free Einstein type equations
are related to a $D-$dimensional theory with sources, just as
in the 5-dimensional case treated in \cite{wpdl92}.

Finally, the last equation in (\r{8}) ($^{(D+1)}R_{\alpha D}= 0$) allows us
to define the $D-$dimensional tensor $P^{\beta}\!_{\alpha}$ such that
this field equation can be expressed as
\be
{P^{\beta}}_{\alpha ; \beta} =0~~ with~~
{P^{\beta}}_{\alpha} = \frac{1}{2\sqrt{g_{DD}}} \left
(g^{\beta \mu} g_{\mu \alpha , D} - {\delta ^\beta}_\alpha
g^{\rho \nu} g_{\rho \nu ,D} \right ).
\ee
As has already been observed in \cite{wpdl92}, these equations have the
appearence of
conservation laws, the full meaning of which is not clear at the present.
%
\section{$D-$dimensional cosmological solutions}
Within the context of the Wesson's scheme (from 5 to 4 dimensions)
discussed in section (1),
a number of
attempts have been made to construct cosmological solutions
in 4 dimensions. In particular, Wesson and Ponce de
Leon \cite{w92,wpdl92} have obtained Friedmann-Lema\^{\i}tre-Robertson-Walker
(FLRW) type solutions with flat spatial sections. These cosmological
solutions essentially fall into two classes, depending upon
whether or not they possess
a Killing vector in their fifth dimension.

In this section we present a generalisation of the more physically
relevant solutions due to Wesson and Ponce de Leon and,
to begin with, we consider the solution in
5 dimensions in the form\footnote{We are grateful to the referee
for pointing out that this solution was first found by
Belinsky and Khalatnikov \cite{bk72}.}
\be
ds^2 = dt^2 -t (dx^2 +dy^2 +dz^2) -t^{-1} d{\psi}^2,
\l{wesson}
\ee
which possesses a Killing vector in its fifth dimension.
This spacetime has two important features: firstly
it has
a shrinking fifth dimension with time $t$, and secondly
the 4-dimensional equations it gives rise to  produce a FLRW model
with a radiative equation of state $p=\frac{1}{3} \rho$,
where
\be
 \rho = \frac{3}{4t^2},~~~~p=\frac{1}{4t^2}.
\ee
Thus in
this sense the the fifth dimension ``generates'' effective sources
(with non-zero $\rho$ and $p$) and these in turn act
to curve the 4-dimensional spacetime.

It is useful here to emphasise a different
interpretation of this scheme by recalling that for
5-dimensional Kaluza-Klein spacetimes
with a symmetry group
generated by $\frac{\partial}{\partial \psi}$
(implying that the
metric is independent of the extra spatial
dimension $\psi$)
 the field equations
\be
^{(5)}{G}_{ij} =0,
\ee
are formally identical to the vacuum Brans-Dicke
field equations in four dimensions with the
free parameter $\omega =0$ \cite{freund82}. Therefore Wesson's
solution (\r{wesson}) should correspond to a vacuum
FLRW solution of the Brans-Dicke theory with $\omega =0$.
This turns out to be the O'Hanlon-Tupper
solution \cite{hanlon72} given by
\be
ds^2 = dt^2 -t^q (dx^2 +dy^2+dz^2),~~~~\phi (t) = \phi_0 t^r,
\l{ohanlon}
\ee
where $\phi$ is the Brans-Dicke scalar field, $\phi_0$ is
a constant and the parameters $r$ and $q$ are given by
\be
\frac{1}{r} =-\frac{1}{2}\left[1\pm \sqrt{3(2\omega +3)}\right ],
{}~q=\frac{1}{3} (1-r).
\ee
Substituting $\omega=0$ into (\r{ohanlon}) gives the Wesson's solution
for $r=- \frac{1}{2}$ (incidentaly, $r=1$ leads to Minkowski spacetime
which by equation (\ref{eq10}) give $T_{\alpha \beta} =0$).
We should note, however, that despite the mathematical
equivalence of the two solutions, the fact that they come from different
physical theories makes them conceptually distinct. Indeed the
O'Hanlon-Tupper solution represents an empty universe, the curvature
of which is generated by a non-static scalar field - usually
related to the Newtonian constant of gravitation through
$\phi \sim G^{-1}$ \cite{brans61}.

Now in order to generalise solution (\r{wesson}) from
$(4+1)$ to $(D+1)$ dimensions, we consider the generalised
spatially flat FLRW line element in the form
\be
ds^2 = dt^2 -R^2 (t) \sum_{i=1}^{D-1} {(d{x^i})^2}
- \phi^2 (t) d\psi^2.
\l{fr}
\ee
The vacuum Einstein field equations in the $(D+1)$
dimensions are given by (8)
which for the metric (\r{fr}) become
\be
(D-1) \frac{\ddot R}{R} + \frac{\ddot \phi}{\phi} =0
\ee
\be
(D-1) \frac{\dot \phi}{\phi}\frac{\dot R}{R}
+\frac{\ddot \phi}{\phi} =0
\ee
\be
\frac{\ddot R}{R} + (D-2)\frac{{\dot R}^2}{R^2}
+\frac{\dot \phi}{\phi}\frac{\dot R}{R}=0.
\ee
The solution of these equations is given by
\be
R(t) = R_0 \left [ \frac{D t}{2} +A\right ]^{(2/D)},
 ~~~~\phi (t) =\frac{B}
{R^{\frac{D}{2}-1} (t)},
\l{22}
\ee
where $R_0, A$ and $B$ are integration constants.
An obvious coordinate transformation then allows the
metric
(\r{fr}) to be put in the following form\footnote{Again, it was brought
to our attention
after submitting this paper that an equivlent solution
had been discovered in another context by Sato \cite{sato84}.}
\be
ds^2 = dt^2 -t^{4/D} \sum_{i=1}^{D-1}(dx^i)^2
-t^{(4/D -2)} d\psi^2,
\ee
which clearly shows that for $D=4$ the Wesson-O'Hanlon-Tupper
metric is recovered. This latter result may be viewed as
a Kasner-like universe in $(D+1)$ dimensions being compactified
to a Friedmann-like universe in $D$ dimensions for an arbitrary
$D>2$. In this sense, the compactification property
of this solution is robust with respect to changes in
dimensions for all $D>2$. Furthermore, it can be easily seen
from the equation (\r{22}) that a comoving observer will
measure a matter density $\rho$ given
by
\be
\rho = \frac{\ddot \phi}{\phi} =\frac{2}{t^2}\left(\frac{1}{D} -1\right)
\left(\frac{2}{D} -1\right)
\ee
and a pressure given by
\be
p = \frac{2}{Dt^2}\left(1-\frac{2}{D}\right).
\ee
which are related by $p={\rho}/({D-1})$ (the general $D-$dependent radiative
equation of state), where $p\rightarrow 0$ as $D$ increases.
\\

Now let us turn our attention to the solutions given by
Ponce de Leon \cite{pdl88}. He considers solutions
which depend on the fifth
coordinate ($\psi$) and by using a method of separation of variables
solves the field equations to give eight families of
solutions. The analogues of these classes of solutions -
in the case of going from four to three dimensions -
will be discussed in the next section.
Here we will consider only two classes of his solutions which
are potentially of more physical interest in the context of $4-$dimensions.
Up to coordinate transformations these solutions are
given by
\be
ds^2 = \frac{\Lambda}{3} \psi^2 dt^2 - \psi^2
e^{2\sqrt{\Lambda /3}t} \left (
dx^2 + dy^2 +dz^2\right ) -d \psi^2
\l{ponce1}
\ee
and
\be
ds^2 = \psi^2 dt^2
-t^{2/\alpha} \psi ^{2/(1-\alpha)}
 \left (
dx^2 + dy^2 +dz^2\right ) - \alpha^2 (1-\alpha)^{-2} t^2 d \psi^2,
\l{ponce2}
\ee
where $\Lambda$ and $\alpha$ are arbitrary parameters.
Now as shown by Ponce de Leon, the solutions (\r{ponce1})
and (\r{ponce2}) may be interpreted in 4 dimensions
(with $\psi =$ constant) as de Sitter
and FLRW metrics respectively. However, we should point out that
although they are claimed to be independent solutions \cite{pdl88},
they in fact represent the same spacetime in five dimensions,
as both have vanishing curvature. Furthermore,
the natural
generalisations of (\r{ponce1})
and (\r{ponce2}) are given by
\be
ds^2 = \frac{\Lambda}{3} \psi^2 dt^2 - \psi^2
e^{2\sqrt{\frac{\Lambda}{3}}t}
\sum_{i=1}^{D-1} {d{x^i}^2} -d\psi^2,
\l{ponce3}
\ee
and
\be
ds^2 = \psi^2 dt^2
-t^{2/\alpha} \psi ^{2/(1-\alpha)}
\sum_{i=1}^{D-1} {d{x^i}^2}
- \alpha^2 (1-\alpha)^{-2} t^2 d \psi^2,
\l{ponce4}
\ee
which are also flat in $(D+1)$ dimensions. In $D$ dimensions the metric
(\r{ponce3})
represents a generalised vacuum de Sitter solution with
a cosmological constant $\Lambda_D =
\frac{\Lambda}{6}(D -1)(D-2)$. So in this way, one could say
that the cosmological constant (as well as matter) can be
viewed as the manifestation of the extra
dimension of spacetime. Similarly, the solution (\r{ponce4}) gives
rise to a $D-$dimensional universe with energy density and pressure given
respectively by
\be
\rho_D = \frac{(D-1)(D-2)}{2\alpha^2 \psi^2 t^2}
,~~~~~p_D = \frac{2(D-2)\alpha - (D-1)(D-2)}{2\alpha^2 \psi^2 t^2},
\ee
which amounts to a perfect fluid equation of state
with $p_D = \lambda_D \rho_D$, where $\lambda_D = \frac{2\alpha}{D-1} -1$
and $\frac{D-1}{2} \le\alpha\le D-1$, resulting in a $D-$dimensional
generalisation of Ponce de Leon's result \cite{pdl88}.
\section{Source free GR and lower gravity theories}
As was discussed in section (1), lower dimensional gravity theories
have in general radically different properties from the usual
4-dimensional gravity. Furthermore, the notion of what
is physical in lower dimensional gravity is not strictly speaking understood.
It is in fact not clear whether such a question can be answered
purely in reference to the lower dimesnional theories themselves.
Therefore, any mechanism that allows a bridge to be established
between such theories and the ordinary 4-D gravity is of vital interest.
One possible way to proceed
is to make sure that the
ansatz employed in  $(2+1)$-dimensional studies have at least
some relation to
the $(3+1)$ Einstein theory, and one way of doing this
 is to start by concentrating on those solutions
in $(2+1)$ dimensions that are related to their vacuum $(3+1)-$dimensional
counterparts via the generalisation of Wesson's procedure
given in section (2). Apart from being
mathematically possible
(shown in section (2)), and having the desirable property
of establishing a
bridge,
we think this is justified as a possible mechanism,
especially in view of the absence (as far as we are aware)
of other physically
more suggestive bridges.
\\

Here, as a first step,
we employ this generalisation
to derive the analogues of the classes of solutions
obtained by Ponce de Leon \cite{pdl88} in $(2+1)$ dimensions. We therefore
similarly start by letting the 4-dimensional vacuum spacetime
to be Ricci-flat, homogenous and isotropic
and for the sake of comparison take as our line element the
four dimensional analogue of the line element proposed by Wesson \cite{w92}
in the form
\be
ds^2 = e^a dt^2 - e^b (dx^2 + dy^2) - e^c d\psi^2,
\ee
where $a$, $b$, and $c$ are differentiable functions of $t$ and $\psi$ only.
The vacuum 4-dimensional field equations are given by
\bea
G_{00}&=&e^{-a}[(b_{,t})^2 + 2b_{,t}c_{,t}] + e^{-c}[-3(b_{,\psi})^2 +
2b_{,\psi}c_{,\psi} - 4b_{,\psi\psi}] = 0
\l{32} \\
G_{03}&=&a_{,\psi}b_{,t} - b_{,\psi}b_{,t} + b_{,\psi}c_{,t} - 2b_{,t\psi} = 0
\l{33} \\
G_{11} = G_{22}&=&e^{-c}[(a_{,\psi})^2 + a_{,\psi}b_{,\psi} - a_{,\psi}
c_{,\psi} + 2a_{,\psi\psi} + (b_{,\psi})^2 - b_{,\psi}c_{,\psi} +
2b_{,\psi\psi}] \nonumber\\
&+&e^{-a}[a_{,t}b_{,t} + a_{,t}c_{,t} - (b_{,t})^2 - b_{,t}c_{,t} - 2b_{,tt}
 - (c_{t})^2 - 2c_{,tt}] = 0
\l{34}\\
G_{33}&=&e^{-c}[2a_{,\psi}b_{,\psi} + (b_{,\psi})^2] + e^{-a}[2a_{,t}b_{,t} -
3(b_{,t})^2 - 4b_{,tt}] = 0.
\l{35}
\eea
To find the solutions of (\r{32})-(\r{35}) we proceed similarly to Ponce de
Leon \cite{pdl88}, and assume that the metric coefficients are separable in
their arguments, thus
\bea
&e^{a}&= N(\psi) T(t)
\l{36} \\
&e^{b}&= P(\psi) S(t)
\l{37} \\
&e^{c}&= M(\psi) K(t),
\l{38}
\eea
where $N,P,M$ are undefined differentiable functions of $\psi$ and
$T,S,K$ are undefined differentiable functions of $t$.
Substitution of equations (\r{36})-(\r{38}) into equation (\r{33}) yields
\be
\frac {S_{,t}}{S}\left( \frac{P_{,\psi}}{P} - \frac{N_{,\psi}}
{N} \right) = \frac{P_{,\psi}}{P} \frac{K_{,t}}{K}
\l{39}.
\ee

The following analogous classes of solutions to those given by
Ponce de Leon \cite{pdl88}
were obtained in $(3+1)$ dimensions. The related 3-dimensional metrics
can then be
obtained from the vacuum 4-dimensional solutions by assuming
$\psi =$ constant.
\\

{\bf (i)} Letting $S$ and  $K$ to be constants, we find the following
relationships between $P(\psi), M(\psi)$ and $P(\psi), N(\psi)$ from
field equations (\r{32}) and (\r{35}),
\be
\frac{({P_{,\psi}})^4}{P} = C_1 M^2, ~~~~
P = \frac {C_2}{N^2}.
\l{39a}
\ee
where $C_1$ and $C_2$ are arbitrary constants.
Rescaling $t$ and $\psi$ according to $T^{\frac {1}{2}} dt = const.
d\bar{t}$ and $M^{\frac {1}{2}} d\psi = const. d\bar{\psi}$, as well as
removing unnecessary constants, the metric
takes the form
\be
ds^2 = {\bar{\psi}}^{-\frac {2}{3}} d{\bar{t}}^2 - {\bar{\psi}}
^{\frac {4}{3}} (dx^2 + dy^2) - d{\bar{\psi}}^2,
\ee
which for $\bar \psi =$ constant
corresponds to the $(2+1)$-dimensional Minkowski metric.
\\

{\bf (ii)} Assuming $P$ and $N$ to be constants  we can again find
relationships between $S, K$ and $T$ of exactly the same functional form
as (\r{39a}). This yields the following metric,
\be
ds^2 = d{\bar{t}}^2 - {\bar{t}}^{\frac {4}{3}} (dx^2 + dy^2)
- {\bar{t}}^{-\frac {2}{3}} d{\bar{\psi}}^2,
\ee
which in $(2+1)$ dimensions is the O'Hanlon-Tupper solution (\r{ohanlon}).
The empty $(3+1)$-dimensional solution has the
interesting property that the fourth coordinate $\bar\psi$ will shrink
relative
to the two space dimensions $x$ and $y$ as $\bar t$ increases. The density
and pressure in $(2+1)$ dimensions are given by $\rho = \frac{4}{9t^2},
p=\frac{2}{9t^2}$, which correspond to
the 3-dimensional radiative equation of state
$p = \frac{\rho}{2}$.
\\

{\bf (iii)} Assuming $S$ and $P$ to be constants and rescaling $t$ and $\psi$
such that $T$ and $M$ become constants, equation (\r{34}) yields the following
relations
\be
2K{\ddot K} - {\dot K}^2 - FK = 0
\l{101}
\ee
and
\be
2N{\ddot N} - {\dot N}^2 - FN = 0
\l{102}
\ee
where $F$ is a constant. For the case $F = 0$ equations (\r{101}) and (\r{102})
can be integrated in terms of elementary functions to generate the
flat metric
\be
ds^2 = {\bar\psi}^2 d{\bar{t}}^2 - (dx^2 + dy^2)
-{\bar{t}}^2 d{\bar{\psi}}^2,
\ee
which for $\bar \psi =$ constant again corresponds to the
$(2+1)$-dimensional
Minkowski metric.
\\

Now if $S_{,t}$ and $P_{,\psi}$ are non-zero, then equation (\r{39}) becomes
\be
\frac {P}{P_{,\psi}}\left(\frac{P_{,\psi}}{P} - \frac{N_{,\psi}}{N}\right)
 = \frac{S}{S_{,t}} \frac{K_{,t}}{K} = \alpha,
\ee
where $\alpha$ is an arbitrary constant. Solving for $N$ and $K$ in terms of
$P$ and
$S$ respectively we can write the metric coefficients as,
\bea
&e^{a}&= C_3 [P(\psi)]^{(1-\alpha)} T(t)
\l{44}\\
&e^{b}&= P(\psi) S(t)
\l{45}\\
&e^{c}&= C_4 M(\psi) [S(t)]^{\alpha},
\l{46}
\eea
where $C_3$ and $C_4$ are arbitrary constants. Substituting (\r{44})-(\r{46})
back into
 the field equation (\r{32}) yields,
\be
\frac {C_4}{C_3} \frac {{S_{,t}}^2 S^{(\alpha - 2)}}{T} (1 + 2\alpha) =
\frac {P^{(1 - \alpha)}}{M} \left(4\frac {P_{,\psi\psi}}{P} - 2\frac
{P_{,\psi}}{P}\frac{M_{,\psi}}{M} -  {\left(\frac {P_{,\psi}}{P}\right)}^2
\right).
\l{51a}
\ee
\\

\noindent Now proceeding analogously to \cite{pdl88},
the following cases arise:
\\

{\bf (iv)} If $\alpha = -\frac{1}{2}$ and we let $P = \psi$ and $T = 1$,
then the metric becomes
\be
ds^2 = C_{3}\bar\psi^{\frac{3}{2}} d{\bar{t}}^2 - {\bar\psi}S (dx^2 + dy^2)
- C_{4}(C_1{\bar\psi}S)^{-\frac{1}{2}} d{\bar{\psi}}^2,
\l{48}
\ee
which upon using using (\ref{35}) gives the following equation for $S(t)$:
\be
4\frac{S_{,tt}}{S^\frac{3}{2}} - \frac{{S_{,t}}^2}{S^\frac{5}{2}} =
\frac{4C_{3}}{C_{4}{C_{1}}^{\frac{1}{2}}}.
\l{49}
\ee
This equation has a solution of the form $S(t) \sim t^{-4}$, with the
constants satisfying $C_3 / (C_4 {C_{1}}^{1/2}) = 16$. Hence,
with suitable reparametrisation, the metric (\r{48}) takes the following
flat form
\be
ds^2 = \bar\psi^{\frac{3}{2}} d{\bar{t}}^2 - {\bar\psi}\bar t^{-4}
(dx^2 + dy^2) - \frac{1}{16}{\bar\psi}^{-\frac{1}{2}}\bar t^2 d{\bar{\psi}}^2.
\ee
This solution has the interesting property
that spatial coordinates $x$ and $y$ contract with
increasing $t$.  The density
and pressure in this case are given by $\rho = \frac{4}{t^2},
p=-\frac{6}{t^2}$, which results in an equation of state that violates both
the strong and dominant energy conditions \cite{hawk}.
This is an example of how
vacuum solutions in $(3+1)$ dimensions
can induce sources in $(2+1)$ dimensions
which according to intuitions derived from
the usual $(3+1)-$dimensional theory would be termed as ``non-physical''.
\\

{\bf (v)} For the case $\alpha \neq -\frac{1}{2}$, equation (\r{51a})
can be split up into the following 2 equations,
\be
\frac {C_4}{C_3} \frac {{S_{,t}}^2 S^{(\alpha - 2)}}{T} (1 + 2\alpha) =
\beta,
\l{51}
\ee
\be
\frac {P^{(1 - \alpha)}}{M} \left(4\frac {P_{,\psi\psi}}{P} - 2\frac
{P_{,\psi}}{P}\frac{M_{,\psi}}{M} -  {\left(\frac {P_{,\psi}}{P}\right)}^2
\right) = \beta,
\l{52}
\ee
where $\beta$ is a nonzero constant. From (\r{51}) we find the following
relationship between $S(t)$ and $T(t)$,
\be
T(t) = \gamma S^{(\alpha - 2)}{S_{,t}}^2
\l{53}
\ee
where
\be
\gamma = \frac{C_4}{C_3} \frac{(1 + 2\alpha)}{\beta},
\ee
while (\r{52}) yields the relation between $P(\psi)$ and $M(\psi)$,
\be
{P_{,\psi}}^2 = \left[ E + \frac{\beta P^{(\alpha + \frac{1}{2})}}{\alpha +
\frac{1}{2}} \right] \frac{MP^{\frac{1}{2}}}{2},
\l{55}
\ee
where E is a constant of integration. Substituting (\r{53}) and (\r{55}) into
(\r{44})-(\r{46}) gives
\bea
&e^a& = \gamma C_3P^{(1-\alpha)}S^{(\alpha-2)}{S_{,t}}^2\\
&e^b& = PS\\
&e^c& = 2C_4 \frac{{P_{,\psi}}^2}{P^{\frac{1}{2}}} {\left[ E +
\frac{\beta P^{\alpha + \frac{1}{2}}}{\alpha + \frac{1}{2}} \right]}^{-1}
S^{\alpha},
\eea
which upon substitution in field equation (\r{35}) yields
the remaining cases as those satisfying the condition
$(2\alpha-3)(1+2\alpha)E = 0$. As we have already considered the case
$\alpha = -\frac{1}{2}$, we shall now consider the cases of $\alpha =
\frac{3}{2}$ and $E = 0$ in turn.
\\

If $\alpha = \frac{3}{2}$ and we choose the coordinate transformation $\gamma
C_3 {S_{,t}}^2 dt^2 = d{\bar t}^2$ and ${P_{,\psi}}^2 d\psi^2 = d{\bar
\psi}^2$, again removing unnecessary constants, then the corresponding
solution becomes
\be
ds^2 = (\bar \psi \bar t)^{-\frac{1}{2}}
d{\bar{t}}^2 - \bar t\bar\psi (dx^2 + dy^2) -
{\bar\psi}^{-\frac{1}{2}}\left[E + {\bar\psi}^2
\right]^{-1}{\bar t}^{\frac{3}{2}} d{\bar{\psi}}^2 .
\ee
This indicates that the coefficient of $d\bar{t}$
decreases as $\bar t$ increases.  The density
and pressure are given by $\rho = \frac{1}{4t^{3/2}},
p=\frac{1}{8t^{3/2}}$, which imply a radiative equation of state
in $(2+1)$ dimensions with $p=\frac{\rho}{2}$.
\\

{\bf (vi) - (viii)} The case where $E = 0$ follows through
exactly as in \cite{pdl88},
thus we are left with the three cases where (a) $\alpha = 0$, (b) $\alpha = 1$
, and (c) $\alpha \neq -\frac{1}{2}, 0, 1, \frac{3}{2}$. The appropriate
solutions (which turn out to be flat) will in each case become:

\be
{\bf (a)}~~~~
ds^2 = C{\bar\psi}^2 d{\bar{t}}^2
- {\bar\psi}^2 exp \left[ 2{\sqrt C}{\bar t} \right] (dx^2 + dy^2)
- d{\bar{\psi}}^2.
\l{eq64}
\ee
This yields the Robertson-Walker-de Sitter metric, where by setting $C =
\Lambda/3$ the scale factor can be written in its usual form of $e^{2\sqrt
{\Lambda /3}}t$. Hence, we recover the $(2+1)-$dimensional equivalent of
equation (\r{ponce3}).

\be
{\bf (b)}~~~~
ds^2 = d{\bar{t}}^2 - {\bar t}^2 exp \left[ 2{\bar\psi}\right] (dx^2 + dy^2)
- {\bar t}^2 d{\bar{\psi}}^2 .
\l{eq65}
\ee
This $(3+1)-$dimensional spacetime leads to a flat $(2+1)$-dimensional
model
with the density
and pressure given by $\rho = \frac{1}{t^2}$
and $ p=0$, corresponding to a dust filled universe.
\be
{\bf (c)}~~~~
ds^2 =  {\bar\psi}^2 d{\bar{t}}^2 -
{\bar t}^{\frac{2}{\alpha}} {\bar\psi}^{\frac{2}{1 - \alpha}} (dx^2 + dy^2)
- \frac{\alpha^2 {\bar{t}}^2}{(1 - \alpha)^2} d{\bar{\psi}}^2,
\l{eq66}
\ee
which yields a $3-$dimensional FLRW metric with scale factor
proportional to $t^{2/\alpha}$, where the density and pressure are
given by $\rho = \frac{1}{\alpha^2t^2}$ and
$p = \frac{1}{t^2}(\frac{1}{\alpha} - \frac{1}{\alpha^2})$. The
resulting equation of state,
$p = (\alpha - 1)\rho$, is the special case of equation
(\r{ponce4}) in $(2+1)$ dimensions.
Comparing our results with those obtained by Ponce de Leon, it can
be seen that solutions (i), (ii), and (iii), although different in details
of the functional forms of their metric coefficients,
are qualitatively the same. This is also true of the solutions
(vi)-(viii) which are in fact identical to those obtained by Ponce de
Leon.  Solutions (iv) and (v) are, however, different. Solution (iv) has
the property that the $x$ and $y$
spatial coordinates shrink with increasing $t$, while in solution (v) the
time coordinate actually  shrinks. It is worthwhile mentioning
here that although the solutions (46), (54), (64), (65) and (66) are
all flat in $(3+1)$ dimensions, i.e. they represent the same
Minkowski spacetime, they give rise (via Wesson's scheme) to
non-equivalent embeddings which generate $(2+1)$ spacetimes
with different extrinsic curvatures.
\\

It is also worth pointing out that Wesson's procedure as employed here
provides a simple
mechanism for generating 3-dimensional spaces which have their symmetries
induced by the symmetries of the 4-dimensional space.
For example, equation
(32) represents a class
of line elements with axial symmetry in four dimensions which
generates a class of circularly symmetric
spacetimes in 3 dimensions. Finally, the examples given above
were meant to act as an illustration of the idea put forward
in this paper. It would be of interest to consider
other embeddings, including further non-flat embedding
spacetimes.
\section{Conclusions}
We have generalised the prodedure put forward by Wesson and co-workers
for setting up a correspondence between vacuum 5-dimensional Kaluza-Klein
theory with 4-dimensional Einstein theory with sources to arbitrary
dimensions, and applied this generalisation to connect the ordinary
vacuum 4-dimensional gravity with $(2+1)$-dimensional
theory with sources. This correspondence between the vacuum  Einstein
theory and lower dimensional theories is of potential importance in view
of the great deal of effort that is currently going into the study of lower
dimensional theories, and particularly in view of the radical
differences between such theories and 4-dimensional gravity.
Furthermore, this correspondence gives a mechanism for generating solutions
in $(2+1)$ dimensions that are naturally related to vacuum
Einstein solutions.

Finally an important question with regards to lower dimensional
theories is how to choose ``reasonable''
forms of the field equations and the energy-momentum tensor.
Our work shows that vacuum Einstein solutions
may not necessarily correspond to solutions in lower dimensions
with reasonable (in the sense of 4-dimensions) equations of state,
which highlights the fact that
intuitions regarding physical reasonableness may not automatically
be transportable to lower dimensions.

Further study of this correspondence is in progress and
will be reported in future.
\\

{\bf Acknowledgements} CR was partially supported by CNPq (Brazil)
and would like to thank the School of Mathematical
Sciences for hospitality. SR was supported by the award of a PPARC
studentship and
RT was supported by SERC UK Grant number H09454.
\newpage


\end{document}